\documentclass{article}
\usepackage{latexsym}
\usepackage{amssymb}
\usepackage{amsmath}
\newtheorem{Theorem}{Theorem}[section]
\newtheorem{lemma}{Lemma}[section]
\newtheorem{prop}{Proposition}[section]
\newtheorem{rem}{Remark}[section]

\begin{document}

\title{Gap Solitons in Periodic Discrete Nonlinear Schr\"odinger Equations}

\author{
{\sc  A. Pankov}\\
Mathematics Department\\
College of William and Mary\\
Williamsburg, VA 23187--8795\\
e-mail: {\tt pankov@member.ams.org} }

\date{}

\maketitle

\begin{abstract}
   It is shown that the periodic DNLS, with cubic nonlinearity, possesses gap solutions, i.~e. standing waves, with
   the frequency in a spectral gap, that are exponentially localized in spatial variable. The proof is based on the
   linking theorem in combination with periodic approximations.\vspace{2ex}

\end{abstract}

\setcounter{section}{-1}
\section{Introduction}

In this paper we consider spatially localized standing waves for the discrete nonlinear Schr\"odinger equation (DNLS)
\begin{equation}\label{0.1}
   i\dot{\psi}_n=-\Delta\psi_n+\varepsilon_n\psi_n-\sigma\chi_n|\psi_n|^2\psi_n,\quad n\in\mathbb{Z},
\end{equation}
where $\sigma=\pm 1$,
   $$ \Delta \psi_n=(\psi_{n+1}+\psi_{n-1}-2\psi_n)$$
is the discrete Laplacian in one spatial dimension and given sequences $\varepsilon_n$ and $\chi_n$ are assumed to
be $N$-periodic in $n$, i.~e. $\varepsilon_{n+N}=\varepsilon_n$ and $\chi_{n+N}=\chi_n$. Such solutions are often
called intrinsic localized modes or breathers, but in the case under consideration we prefer the name ''gap
solitons`` due to the obvious analogy with gap solitons in photonic crystals (see, e.~g.
\cite{Aceves,BronskiSeWe,Mills}).

Making use the standing wave Ansatz
   $$ \psi_n=u_n\exp(i\omega t),$$
where $u_n$ is a real valued sequence and $\omega\in\mathbb{R}$, we arrive at the equation
\begin{equation}\label{0.2}
   -\Delta u_n+\varepsilon_nu_n-\omega u_n=\sigma\chi_n|u_n|^2u_n.
\end{equation}
We impose the following boundary condition at infinity:
\begin{equation}\label{0.3}
   \lim_{n\to\pm\infty}u_n=0,
\end{equation}
and we are looking for nontrivial solutions, i.~e. solutions that are not equal to 0 identically.

Actually, we consider a more general equation
\begin{equation}\label{0.4}
   Lu_n-\omega u_n=\sigma\chi_n|u_n|^2u_n
\end{equation}
with the same boundary condition (\ref{0.3}). Here $L$ is a second order difference operator
   $$ Lu_n=a_nu_{n+1}+a_{n-1}u_{n-1}+b_nu_n$$
where $a_n$ and $b_n$ are real valued $N$-periodic sequences. The operator $L$ can be represented in the form
   $$ Lu_n=-(\partial^*a_n\partial)f_n+(a_{n-1}+a_n+b_n)u_n,$$
where
   $$ \partial u_n=u_{n+1}-u_n,\quad \partial^*u_n=u_{n-1}-u_n.$$
When $a_n\equiv 1$ and $b_n=-2+\varepsilon_n$, we obtain equation (\ref{0.2}).

We consider equation (\ref{0.4}) as a nonlinear equation in the space $l^2$ of two-sided infinite sequences. Note that
every element of $l^2$ automatically satisfies (\ref{0.3}).

The operator $L$ is a bounded and self-adjoint operator in $l^2$. Its spectrum $\sigma(L)$ has a band structure, i.~e.
$\sigma(L)$ is a union of a finite number of closed intervals (see, e.~g., \cite{Teschl}). The complement
$\mathbb{R}\setminus\sigma(L)$ consists of a finite number of open intervals called spectral gaps. Two of them are
semi-infinite. We fix one such gap and denote it by $(\alpha,\beta)$.

Our main result is the following
\begin{Theorem}\label{t.1}
   Suppose that $\chi_n>0$ and $\omega\in(\alpha,\beta)$. If either $\sigma=+1$ and $\beta\ne+\infty$, or $\sigma=-1$
   and $\alpha\ne-\infty$, then equation (\ref{0.4}) has a nontrivial solution $u\in l^2$ and, moreover, the solution
   $u$ decays exponentially at infinity:
      $$ |u_n|\le Ce^{-\gamma|n|}, \quad n\in\mathbb{Z},$$
   with some $C>0$ and $\gamma>0$. If either $\sigma=+1$ and $\beta=+\infty$, or $\sigma=-1$ and $\alpha=-\infty$,
   then there is no nontrivial solution in $l^2$.
\end{Theorem}

The proof contained in Sections~\ref{S.1}--\ref{S.5} is variational. Its idea is borrowed from \cite{Pankov} and
based on so-called periodic approximations. In what follows we consider the case $\sigma=+1$. The other case
reduces to the previous one if we replace $L$ by $-L$ and $\omega$ by $-\omega$.

\section{Variational setting}\label{S.1}
\setcounter{equation}{0} \setcounter{Theorem}{0} \setcounter{lemma}{0} \setcounter{prop}{0}

On the Hilbert space $E=l^2$, we consider the functional
\begin{equation}\label{1.1}
   J(u)=\frac 12\,(Lu-\omega u, u)-\frac 14\sum_{n=-\infty}^{+\infty}\chi_n u_n^4,
\end{equation}
where $(\cdot,\cdot)$ is $l^2$ inner product. The corresponding norm in $E$ is denoted by $\|\cdot\|$. The
functional $J$ is a well-defined $C^1$ functional on $E$ and equation (\ref{0.4}) is easily recognized as the
corresponding Euler-Lagrange equation for $J$ (remind that $\sigma=+1$). Thus, we are looking for nonzero critical
points of $J$.

Fix an integer $k>0$ and denote by $E_k$ the space of all $kN$-periodic sequences. This is a $kN$-dimensional Hilbert
space endowed with the inner product
   $$ (u,v)_k=\sum_{n=0}^{kN-1}u_nv_n,\quad u,v\in E_k,$$
and corresponding norm $\|\cdot\|_k$. On the space $E_k$ we consider the functional
\begin{equation}\label{1.2}
   J_k(u)=\frac 12\,(Lu-\omega u, u)_k-\frac 14\sum_{n=0}^{kN-1}\chi_n u_n^4.
\end{equation}
Due to the periodicity of coefficients the operator $L$ acts in $E_k$. Critical points of $J_k$ are exactly
$kN$-periodic solutions of equation (\ref{0.4}) with $\sigma=+1$.

For gradients of $J$ and $J_k$ we have the following formulas:
\begin{equation}\label{1.3}
   \big(\nabla J(u),v\big)=(Lu-\omega u,v)-\sum_{n=-\infty}^{+\infty}\chi_nu_n^3v_n,\quad v\in E,
\end{equation}
and
\begin{equation}\label{1.4}
   \big(\nabla J_k(u),v\big)=(L_ku-\omega u,v)_k-\sum_{n=0}^{kN-1}\chi_nu_n^3v_n,\quad v\in E.
\end{equation}

We denote by $L_k$ the operator $L$ acting in $E_k$. From the spectral theory of difference (Jacobi) operators (see,
e.~g., \cite{Teschl}) it follows immediately that $\sigma(L_k)\subset\sigma(L)$ and, hence, $\|L_k\|\le\|L\|$.

Let $E_k^+$ (respectively, $E_k^-$) be the positive (respectively, negative) spectral subspace of the operator
$L_k-\omega$ in $E_k$. Similarly, we introduce the positive and negative spectral subspaces $E^+\subset E$ and
$E^-\subset E$, respectively, for the operator $L-\omega$. Let
   $$ \delta=\min\big[|\alpha-\omega|,|\beta-\omega|\big]$$
be the distance from $\omega$ to the spectrum $\sigma(L)$. Then
\begin{equation}\label{1.5}
   \pm(Lu-\omega u,u)\ge\delta\|u\|^2,\quad u\in E^\pm,
\end{equation}
and
\begin{equation}\label{1.6}
   \pm(L_ku-\omega u,u)_k\ge\delta\|u\|^2_k,\quad u\in E_k^\pm.
\end{equation}

Now we are ready to prove the nonexistence part of Theorem~\ref{t.1}.

\begin{prop}\label{p.1}
   Suppose that $\beta=+\infty$. Then the only critical point of $J$ (respectively, $J_k$) is the origin of the space
   $E$ (respectively, $E_k$).
\end{prop}

{\bf Proof.} We consider the case of $J$, the remaining case being similar.

Let $u\in E$ be a critical point of $J$. Then, by (\ref{1.3}) and positivity of $\chi_n$,
\begin{eqnarray*}
   0 &=& \big(\nabla J(u),u\big)=(Lu-\omega u,u)-\sum_{n=-\infty}^{+\infty}\chi_nu_n^4\le\\
     &\le & (Lu-\omega u,u).
\end{eqnarray*}
Since  $\beta=+\infty$, we have that $E^+=\{0\}$ and, by (\ref{1.5}),
   $$ 0\le -\delta\|u\|^2$$
which implies that $u=0$.\hfill$\Box$

\section{Technical results}\label{S2}
\setcounter{equation}{0} \setcounter{Theorem}{0} \setcounter{lemma}{0} \setcounter{prop}{0}

To prove the existence of $kN$-periodic solutions, as well as to pass to the limit as $k\to\infty$, we need some
preliminaries. We start with

\begin{lemma}\label{l.1}
   For any nontrivial critical points $u^{(k)}\in E_k$
   of $J_k$ and $u\in E$ of $J$, with critical values $c^{(k)}=J_k(u^{(k)})$ and $c=J(u)$, we have
      $$ \big\|u^{(k)}\big\|_k\le 4\delta^{-1}\overline{\kappa}\underline{\kappa}^{-3/4}(c^{(k)})^{3/4}$$
   and
      $$ \|u\|\le 4\delta^{-1}\overline{\kappa}\underline{\kappa}^{-3/4}c^{3/4},$$
   where $\underline{\kappa}=\min\{\chi_n\}$ and $\overline{\kappa}=\max\{\chi_n\}$.
\end{lemma}

{\bf Proof.} We have
\begin{eqnarray}\label{2.1}
   c &=& J(u)-\frac 12\,\big(\nabla J(u),u\big)=
         \left(\frac 12-\frac 14\right)\sum_{n=-\infty}^{+\infty}\chi_nu_n^4\ge\nonumber\\
     &\ge & \frac 14\,\underline{\kappa}\,\|u\|^4_{l^4},
\end{eqnarray}
where $\|\cdot\|_{l^p}$ stands for the norm in the space $l^p$. Let $u^\pm$ be the orthogonal projection of $u$
into $E^\pm$ along $E^\mp$. Then
\begin{eqnarray*}
   0 &=& \big(\nabla J(u),u^+\big)=(Lu-\omega u, u^+)-\sum_{n=-\infty}^{+\infty}\chi_nu_n^3u_n^+=\\
     &=& (Lu^+-\omega u^+,u^+)-\sum_{n=-\infty}^{+\infty}\chi_nu_n^3u_n^+
\end{eqnarray*}
and we obtain, using H\"older's inequality, that
\begin{eqnarray*}
   \delta\|u^+\|^2 &\le & \overline{\kappa}\left[\sum_{n=-\infty}^{+\infty}u_n^6\right]^{1/2}
                        \left[\sum_{n=-\infty}^{+\infty}(u_n^+)^2\right]^{1/2}=\\
                 &=& \overline{\kappa}\|u\|^3_{l^6}\|u^+\|\le \overline{\kappa}\|u\|^3_{l^4}\|u^+\|.
\end{eqnarray*}
Hence, by (\ref{2.1}),
   $$ \|u^+\|^2\le 2^{3/2}\delta^{-1}\overline{\kappa}\underline{\kappa}^{-3/4}c^{3/4}\|u^+\|.$$
Similarly,
   $$ \|u^-\|^2\le 2^{3/2}\delta^{-1}\overline{\kappa}\underline{\kappa}^{-3/4}c^{3/4}\|u^-\|.$$
Since
   $$ \|u\|^2=\|u^+\|^2+\|u^-\|^2$$
and
   $$ \|u^+\|+\|u^-\|\le 2^{1/2}\|u\|,$$
the two previous inequalities imply that
   $$ \|u\|\le 4\delta^{-1}\overline{\kappa}\underline{\kappa}^{-3/4}c^{3/4}.$$

The remaining case of $J_k$ is similar.\hfill$\Box$\vspace{2ex}

We need also lower estimates for nontrivial critical points and critical values.

\begin{lemma}\label{l.2}
   Under the notation of Lemma~\ref{l.1}
      $$ \|u\|^2\ge 2^{-1/2}\delta\overline{\kappa}^{-1},$$
      $$ c\ge \frac 18\,\delta^2\overline{\kappa}^{-2}\underline{\kappa},$$
      $$ \big\|u^{(k)}\big\|_k^2\ge 2^{-1/2}\delta\overline{\kappa}^{-1}$$
   and
      $$ c^{(k)}\ge\frac 18\,\delta^2\overline{\kappa}^{-2}\underline{\kappa}.$$
\end{lemma}

{\bf Proof.} Consider the case of $J$, the other case being similar.

Since $J'(u)=0$, then $\big(J'(u),v\big)=0$ for any $v\in E$. Taking $v=u^+$, the orthogonal projection of $u$ onto
$E^+$, we have as in the proof of Lemma~\ref{l.1}
   $$ \delta\|u^+\|^2\le\sum_{n\in\mathbb{Z}}\chi_nu^3_nu_n^+.$$
Using the H\"older inequality with $p=4/3$ and $p'=4$, we obtain
   $$ \delta\|u^+\|\le\overline{\kappa}\,\|u\|^3_{l^4}\|u^+\|_{l^4}\le
        \overline{\kappa}\,\|u\|^3\|u^+\|.$$
Similarly, taking $v=u^-$, the orthogonal projection of $u$ onto $E^-$, we obtain that
   $$ \delta\|u^-\|^2\le\overline{\kappa}\,\|u\|^3\|u^-\|.$$
Combining the last two inequalities, we have that
   $$ \delta\|u\|^2\le\overline{\kappa}\,\|u\|^3\left(\|u^+\|+\|u^-\|\right)\le
       2^{1/2}\overline{\kappa}\,\|u\|^4.$$
Hence
   $$ \|u\|^2\ge 2^{-1/2}\delta\overline{\kappa}^{-1}.$$

The bound for $c$ follows immediately from the last inequality and Lemma~\ref{l.1}.\hfill$\Box$

\section{Existence of periodic solutions}
\setcounter{equation}{0} \setcounter{Theorem}{0} \setcounter{lemma}{0} \setcounter{prop}{0}

In this section we prove the existence of nontrivial $kN$-periodic solutions to equation (\ref{0.4}). Actually, we
show that the functional $J_k$ possesses a nontrivial critical point. Moreover, we derive some uniform in $k$
bounds for the critical points and corresponding critical values. The proof relies upon the standard linking
theorem (see Appendix~B).

Recall that a sequence $v^{(j)}\in E_k$ is called a {\it Palais-Smale sequence} for $J_k$ at level $b$ if
$J_k(v^{(j)})\to b$ and $J'_k(v^{(j)})\to 0$ as $j\to\infty$.

\begin{lemma}\label{l.3}
   The functional $J_k$ satisfies the so-called Palais-Smale condition, i.~e. every Palais-Smale sequence contains
   a convergent subsequence.
\end{lemma}

{\bf Proof.} Since the space $E_k$ is finite dimensional, it is enough to show that every Palais-Smale sequence is
bounded. Moreover, replacing $L$ by $L+\omega_0$ and $\omega$ by $\omega+\omega_0$, with some $\omega_0$, we can
assume that $L\gg 1$, i.~e.
   $$ (Lv,v)_k\ge \|v\|_k^2,\quad v\in E_k,$$
and $\omega>0$.

Let $v^{(j)}$ be a Palais-Smale sequence at some level $b$. Fix $\beta\in\left(\displaystyle\frac 14,\frac
12\right)$. For $j$ large, we have
\begin{eqnarray*}
   b+1+\beta\big\|v^{(j)}\big\|_k &\ge & J_k(v^{(j)})-\beta\left(J'_k(v^{(j)}),v^{(j)}\right)=\\
        &=& \left(\frac 12-\beta\right)\left(Lv^{(j)},v^{(j)}\right)_k-
            \left(\frac 12-\beta\right)\omega\big\|v^{(j)}\big\|_k^2+\\
        & & {}+ \left(\beta-\frac 14\right)\sum_{n=0}^k\chi_n\big(v^{(j)}\big)^4\ge\\
        &\ge & \left(\frac 12-\beta\right)\big\|v^{(j)}\big\|_k^2-
               \left(\frac 12-\beta\right)\omega\big\|v^{(j)}\big\|^2_k+\\
        & & {}+ \left(\beta-\frac 14\right)\underline{\kappa}\,\big\|v^{(j)}\big\|^4_k.
\end{eqnarray*}
Since $a^2\le K(\varepsilon)+\varepsilon a^4$, where $K(\varepsilon)\to\infty$ as $\varepsilon\to 0$, and
$\omega>0$, we can choose $\varepsilon$ so small that the third term above absorbs the second one up to a constant.
Hence
   $$ b+1+\beta\big\|v^{(j)}\big\|_k\ge\left(\frac 12-\beta\right)\big\|v^{(j)}\big\|_k^2+C\big\|v^{(j)}\big\|^4_k-
      C_0,$$
with $C>0$ and $C_0>0$. This implies immediately that the sequence $\big\|v^{(j)}\big\|_k$ is bounded and the proof
is complete.\hfill$\Box$\vspace{2ex}

Now let us check that the functional $J_k$ possesses the linking geometry (see Appendix~\ref{A.B}), with $Y=E^-_k$
and $Z=E^+_k$. Remind that we consider the case when $\beta\ne+\infty$ and, hence, $E^+_k\ne\{0\}$. Fix two
constants $\varrho>r>0$ and choose $z^k\in E^+_k$ as follows.

Let $z\in E^+$ be an arbitrary unit vector. We set
   $$ z^k=\frac 1{\|P^+_kS_kz\|_k}\,P_k^+S_kz\in E^+_k$$
(see Appendix~\ref{A.A} for the definition of operator $S_k$). The vector $z^k$ is well-defined at least for
sufficiently large $k$, say, $k\ge k_0$. If $k<k_0$, we choose $z^k$ to be an arbitrary unit vector in $E^+_k$. Now
we set
   $$ S=\left\{v\in E^+_k \;:\; \|v\|_k=r\right\}$$
and
   $$ M=\left\{v=y+tz^k \;:\; \|v\|_k\le\varrho,\, t\ge 0,\, y\in E^-_k\right\}.$$
Let $M_0$ be the boundary of the set $M$,
   $$ M_0=\left\{v=y+tz^k\;:\; y\in E^-_k,\; \|v\|_k=\varrho  \text{ and } t\ge 0,\text{ or }
                  \|y\|_k\le\varrho \text{ and } t=0\right\}.$$

\begin{lemma}\label{l.4}
   For $\varrho>0$ large enough
      $$J_k(v)\le 0,\quad v\in M_0,$$
   while
      $$ J_k(v)\ge\frac \delta 4\, r^2,$$
   provided $r^2\le \overline{\kappa}^{-1}\delta $. Moreover, there exists a constant $C>0$ independent of $k$ and such
   that
   \begin{equation}\label{3.0}
      J_k(v)\le C,\quad v\in M.
   \end{equation}
\end{lemma}

{\bf Proof.} For $v\in E^+_k$ we have
\begin{eqnarray*}
   J_k(v) &=& \frac 12\,(L_kv-\omega v, v)_k-\frac 14\sum_{n=0}^{kN-1}\chi_nv_n^4\ge\\
          &\ge & \frac\delta 2\,\|v\|^2_k-\frac{\overline{\kappa}}4\,\|v\|_k^4.
\end{eqnarray*}
This implies that, for $r^2\le(\overline{\kappa})^{-1}\delta$,
   $$ J_k\ge\frac\delta 4\,r^2\quad \text{ on } S.$$

Now let us consider $J_k$ on $M$. Since $E^\pm_k$ are mutually orthogonal spectral subspaces of $L_k$,
\begin{eqnarray*}
   J_k(y+tz) &=& \frac 12\,(L_ky-\omega y,y)_k+\frac{t^2}2\,(L_kz^k-\omega z^k,z^k)_k-\\
             & &{} -\frac 14\sum_{n=0}^{kN-1}\chi_n(y_n+tz^k_n)^4.
\end{eqnarray*}
Hence,
   $$ J_k(y+tz)\le-\frac\delta 2\,\|y\|^2_k+\frac{t^2}2\big((L_k-\omega)z^k,z^k\big)_k-
            \frac{1}4\,\underline{\kappa}\,\|y+tz^k\|^4_{l^4_k}.$$

Consider the subspace $X=E^-_k\oplus\mathbb{R}z^k\subset E_k$ endowed with the norm $\|\cdot\|_{l^4_k}$ (for the
definition of $\|\cdot\|_{l^p_k}$ see Appendix~A). The map $y+tz^k\mapsto tz^k$ is a bounded projector onto
$\mathbb{R}z$. Since its norm is not less that 1, we see that
   $$ \|y+tz^k\|_{l^4_k}\ge \|tz^k\|_{l^4_k}.$$
Hence,
\begin{eqnarray}\label{3.1}
   J_k(y+tz^k)\le -\frac\delta 2\,\|y\|^2_k+\frac{t^2}2\big((L_k-\omega)z^k,z^k\big)_k-
                       \frac 14\,\underline{\kappa}t^4\|z^k\|^4_{l^4_k}.
\end{eqnarray}

We have that
   $$ \big((L_k-\omega)z^k,z^k\big)\le a_0,$$
where $a_0=\|L_k-\omega\|$, and, by (\ref{A.2}) and Lemma~\ref{l.A.1},
   $$ \lim_{k\to\infty}\|z^k\|^4_{l^4_k}=\|P^+z\|^4_{l^4}=\|z\|^4_{l^4}.$$
Inequality (\ref{3.1}) implies that
\begin{eqnarray}\label{3.1.a}
   J_k(y+tz^k)\le -\frac\delta 2\,\|y\|^2_k+\frac{a_0}2\,t^2-a_1t^4\le \frac{a_0}2\,t^2-a_1t^4,
\end{eqnarray}
with some $a_1>0$ independent of $k$. Therefore, for all $\varrho$ large we have that $J_k\le 0$ on $M_0$.
 Moreover,
   $$ \sup_MJ_k(v)\le C=\max_{t>0}\left(\frac{a_0}2\,t^2-c_1\,t^4\right),$$
with $C>0$ independent of $k$.  The proof is complete.\hfill$\Box$

Now we are ready to prove the existence of periodic solutions. Remind that, without loss of generality, we consider
the case $\sigma=+1$.

\begin{Theorem}\label{t.3.1}
   Suppose that $\chi_n>0$ and $\omega\in(\alpha,\beta)$, with $\beta\ne+\infty$. Then for every $k\ge 1$ equation
   (\ref{0.4}), with $\sigma=+1$, has a nontrivial $kN$-periodic solution $u^{(k)}$. Moreover, we have the
   following bounds:
      $$ J_k(u^{(k)})\le C,$$
      $$ \|u^{(k)}\|_k\le C, $$
   where $C>0$ is independent of $k$.
\end{Theorem}

{\bf Proof.} The existence follows immediately from the standard linking theorem (see Appendix~\ref{A.B}). Indeed,
to apply that theorem we need two things: $(a)$ the Palais-Smale condition and $(b)$ the linking geometry. The
Palais-Smale condition is verified in Lemma~\ref{l.3}, while Lemma~\ref{l.4} means exactly that the functional
$J_k$ possesses the linking geometry.

Moreover, the linking theorem states that the corresponding critical value satisfies
   $$ J_k\big(u^{(k)}\big)\le\sup_MJ_k(v).$$
Therefore, the upper bounds for the critical points and critical values follow from Lemma~\ref{l.4} and
Lemma~\ref{l.1}.\hfill$\Box$

\section{Existence of localized solutions}\label{S.4}
\setcounter{equation}{0} \setcounter{Theorem}{0} \setcounter{lemma}{0} \setcounter{prop}{0}

The following result gives the existence of solution $u\in l^2$ in Theorem~\ref{t.1}.

\begin{Theorem}\label{t.4.1}
   Under assumptions of Theorem~\ref{t.3.1} equation (\ref{0.4}) possesses a nontrivial solution $u\in l^2$.
\end{Theorem}

{\bf Proof.} Consider the sequence $u^{(k)}=\{u_n^{(k)}\}$ of $kN$-periodic solutions found in Theorem~\ref{t.3.1}.

First we claim that there exists $\delta_0>0$ and $n_k\in\mathbb{Z}$ such that
\begin{equation}\label{4.1}
   \big|u^{(k)}_{n_k}\big|\ge\delta_0.
\end{equation}
Indeed, if not, then $u^{(k)}\to 0$ in $l^\infty$. Hence, $v^{(k)}=R_ku^{(k)}\to 0$ in $l^\infty$. By
Theorem~\ref{3.1}, $\|v^{(k)}\|_{l^2}=\|u^{(k)}\|_k$ is bounded. Now the following simple inequality
   $$ \|v\|^p_{l^p}\le \|v\|^{p-2}_{l^\infty}\|v\|^2_{l^2},$$
where $p>2$, shows that $v^{(k)}\to 0$ in all $l^p$, $p>2$. Hence, $\|u^{(k)}\|_{l^p_k}\to 0$ for all $p>2$. As in
the beginning of proof of Lemma~\ref{l.1}, we have that for the corresponding critical value $c^k=J_k(u^{(k)})$
   $$ 0<c^k=\frac 14\sum_{n\in Q_k}\chi_n\big[u_n^{(k)}\big]^4\le
                         \frac{\overline{\kappa}}4\,\|u^{(k)}\|^4_{l^4_k}\to 0.$$
However, this contradicts Lemma~\ref{l.2} and we obtain (\ref{4.1}).

Due to periodicity of coefficients, $\{u^{(k)}_{n+N}\}$ is also a solution of (\ref{0.4}). Hence, making such
shifts, we can assume that $0\le n_k\le N-1$ in (\ref{4.1}). Moreover, passing to a subsequence of $k$'s, we can
even assume that $n_k=n_0$ is independent of $k$.

Next we extract a subsequence, still denoted by $u^{(k)}$, such that $u^{(k)}_n\to u_n$ for every $n\in
\mathbb{Z}$. Inequality (\ref{4.1}) implies that $|u_{n_0}|\ge\delta_0$ and, hence, $u=\{u_n\}$ is a nonzero
sequence. It is not difficult to show that equation (\ref{0.4}) possesses point-wise limits and, therefore, $u$ is
a solution of that equation.

Finally we have that
   $$ \sum_{n=-a}^a\big|u^{(k)}_n\big|^2\le\|u^{(k)}\|_k\le C$$
for any fixed $a\in\mathbb{Z}$ and $k$ large enough. Passing to the limit, we have that
   $$ \sum_{n=-a}^a|u_n|^2\le C.$$
Since $a$ is arbitrary, $u\in l^2$.\hfill$\Box$

\begin{rem}\label{r.4.1}\rm
   Consider equation (\ref{0.4}) with small nonlinearity, i.~e.
      $$ Lu_n-\omega u_n=\sigma\lambda\chi_n|u_n|^2u_n,$$
   where $\lambda>0$ is a small parameter (all other data, including $\omega$, are fixed). Then Lemma~\ref{l.1}
   shows that for the solution $u=u_\lambda\in l^2$ obtained in Theorem~\ref{t.1} we have that
      $$ \|u_\lambda\|^2\ge c\lambda^{-1},$$
   with $c>0$. This means that the solution $u_\lambda$ bifurcates from infinity. Actually, one can scale out
   $\lambda$ and obtain the same conclusion directly.
\end{rem}

\section{Exponential decay and nonexistence result}\label{S.5}
\setcounter{equation}{0} \setcounter{Theorem}{0} \setcounter{lemma}{0} \setcounter{prop}{0}

To complete the proof of Theorem~\ref{t.1}, we have to show that the solution obtained decays exponentially fast.
Actually, we have

\begin{Theorem}\label{t.5.1}
   Under assumptions of Theorem~\ref{t.3.1}, let $u\in l^2$ be a solution of equation (\ref{0.4}). Then $u$
   satisfies
      $$ |u_n|\le Ce^{-\gamma|n|},\quad n\in\mathbb{Z},$$
   with some $C>0$ and $\gamma>0$.
\end{Theorem}

{\bf Proof.} Let $v_n=-\sigma\chi_n|u_n|^2$. Then
\begin{equation}\label{5.1}
   \widetilde{L}u_n-\omega u_n=0,
\end{equation}
where
   $$ \widetilde{L}u_n=Lu_n+v_nu_n.$$
Since $\lim_{|n|\to\infty}v_n=0$, the multiplication by $v_n$ is a compact operator in $l^2$. Hence,
   $$\sigma_{ess}(\widetilde{L})=\sigma_{ess}(L),$$
where $\sigma_{ess}$ stands for the essential spectrum. Now (\ref{5.1}) means that $u=\{u_n\}$ in an eigenfunction
that corresponds to the eigenvalue of finite multiplicity $\omega\notin\sigma_{ess}(\widetilde{L})$ of the operator
$\widetilde{L}$. Therefore, the result follows from the standard theorem on exponential decay for such
eigenfunctions (see, e.~g. \cite{Teschl}).\hfill$\Box$

The following result shows that if $\omega\in\sigma(L)$, then equation (\ref{0.4}) has no well-decaying (e.~g.,
exponentially fast) nontrivial solution.

\begin{Theorem}\label{t.5.2}
   Suppose that $\omega\in\sigma(L)$ and $u$ is a solution of (\ref{0.4}) such that $|n|^{1/2}u_n\in l^2$. The
   $u_n\equiv 0$.
\end{Theorem}

{\bf Proof.} We have that $u$ satisfies (\ref{5.1}) and this means that $\omega$ is an eigenvalue embedded into
$\sigma_{ess}(\widetilde{L})$. The potential $v=\{v_n\}$ satisfies $|n|v_n\in l^1$ and, by Theorem~7.11,
\cite{Teschl}, $\sigma_{ess}(L)$ is absolutely continuous, hence, contains no embedded eigenvalues.\hfill$\Box$

\section{An extension of main result}\label{S.6}
\setcounter{equation}{0} \setcounter{Theorem}{0} \setcounter{lemma}{0} \setcounter{prop}{0}

Consider the following equation
\begin{equation}\label{6.1}
   Lu_n-\omega u_n=\sigma f_n(u_n)
\end{equation}
which is more general that (\ref{0.4}). Here the operator $L$ is of the same form as above, $\omega$ belongs to
some spectral gap $(\alpha,\beta)$ of $L$ and  $\sigma=\pm 1$. The nonlinearity $f_n(u)$ is supposed to satisfy the
following assumptions.
\begin{description}
   \item[$(i)$] {\it The function} $f_n(u)$ {\it is continuous in} $u\in \mathbb{R}$ {\it and depends periodically
                in } $n$, {\it with period} $N$.
   \item[$(ii)$] {\it There exist } $p>2$ {\it and} $c>0$ {\it such that}
                     $$ 0\le f_n(u)\le c|u|^{p-1}$$
                 {\it near} $u=0$.
   \item[$(iii)$] {\it There exists} $\mu>2$ {\it such that}
                     $$ 0<\mu F_n(u)\equiv \int_0^uf_n(t)\,dt\le f_n(u)u,\quad u\ne 0.$$
\end{description}

Arguing as in the proof of Theorem~\ref{t.1}, with corresponding modifications (see \cite{Pankov} for a similar
result for continuum periodic nonlinear Schr\"odinger equations), we obtain the following result.

\begin{Theorem}\label{t.6.1}
   Under assumptions $(i)$--$(iii)$ suppose that either $\sigma=+1$ and $\beta\ne +\infty$, or $\sigma=-1$ and
   $\alpha\ne -\infty$. Then equation (\ref{6.1}) has a nontrivial  exponentially decaying solution. If either
   $\sigma=+1$ and $\beta=+\infty$, or $\sigma=-1$ and $\alpha=-\infty$, then there is no nontrivial solution in
   $l^2$.
\end{Theorem}

The most important example is the power nonlinearity
\begin{equation}\label{6.2}
   f_n(u)=\chi_n|u|^{p-2}u,
\end{equation}
with $p>2$. If $p=4$, we obtain the cubic nonlinearity considered in Theorem~\ref{t.1}

Theorem~\ref{t.6.1} can be extended immediately to the case of equation (\ref{6.1}) on $\mathbb{Z}^d$, $d\ge 1$,
where
   $$ Lu_n=\sum_{m\in\mathbb{Z}^d}a(n,m)\,u_m.$$
Here $a(n,m)$ satisfies
   $$ a(n+N,m+N)=a(n,m)$$
for some $N=(N_1,N_2,\dots,N_d)$ and
   $$ a(n,m)=0$$
whenever $|n-m|\ge a_0>0$. The nonlinearity $f_n$, $n\in\mathbb{Z}^d$, must satisfy $(i)$--$(iii)$ and the
periodicity assumption
   $$ f_{n+N}(u)=f_n(u),$$
with the same $N$ as above.

\section{Concluding remarks}\label{S.7}
\setcounter{equation}{0} \setcounter{Theorem}{0} \setcounter{lemma}{0} \setcounter{prop}{0}

In the past decade, localized solutions of DNLS has become a topic of intense research. Much of this work concerns
the standard constant coefficient cubic DNLS  and has been summarized in reviews
\cite{FlashWi,HennigTs,KevrekidesRaBi}. Constant coefficient DNLS with general power nonlinearity (\ref{6.2})
($\chi_n\equiv 1$) is considered in \cite{Weinstein}.

Certainly, DNLS with periodic coefficients is not less important. In this case a new phenomenon appears. While the
spectrum of $-\Delta$ consists of a single closed interval, in the spectrum of periodic operator $L$ finite gaps
typically open up. The corresponding DNLS may have standing wave solutions with carrier  frequency in such a gap.
Theorems~\ref{t.1} and \ref{t.6.1} give rigorous results of this type.

Note that in the case of periodic DNLS, and even for more general equations, with $\omega$ below or above the
spectrum (depending on $\sigma$), such solutions are shown to exist in \cite{PankovZa}. In that paper the so-called
Nehari manifold approach is employed, together with a discrete version of concentration compactness principle. For
such values of $\omega$ the quadratic part of the corresponding functional is positive or negative definite. This
fact simplifies the situation considerably. When $\omega$ lies in a finite gap, the quadratic part of the
functional is strictly indefinite. This suggests us to use the linking theorem combined with periodic
approximations. Such approach was used before in
\cite{Pankov05,Pankov,PankovPf00,PankovPf98,PankovPf99,Rabinowitz91}. Another approach to localized solutions of
periodic DNLS, based on the generalized linking theorem (see, e.~g. \cite{Willem}) will be discussed elsewhere.

Note that two-dimensional discrete gap solitons are observed experimentally \cite{FleischerCar,FleischerSeg}. In
\cite{GorbachJoh} results on numerical simulation of gap solutions in a particular periodic DNLS are
reported.\footnote{The author thanks J. Fleischer and A. Gorbach for these references.}

\appendix
\section{Operator $L$}\label{A.A}
\setcounter{equation}{0} \setcounter{Theorem}{0} \setcounter{lemma}{0} \setcounter{prop}{0}

Let $l$ denote the vector spaces of all two sided complex valued sequences $u=\{u_n\}_{n\in\mathbb{Z}}$ and
$l^p\subset l$, $1\le p\le \infty$, the subspace of all $p$-summable (bounded  if $p=\infty$) sequences. Endowed
with the standard norm $\|\cdot\|_{l^p}$, $l^p$ is a Banach spaces (Hilbert space when $p=2$).

Let $N\ge 1$ be a given integer and $a_n$ and $b_n$ two real valued sequences. The formula
\begin{equation}\label{A.1}
   Lu_n=a_mu_{n+1}+a_{n-1}u_{n-1}+b_nu_n
\end{equation}
defines a linear operator acting in the space $l$. Moreover, $L$ is a bounded linear operator in the space $l^p$,
$1\le p\le\infty$, and this is a self-adjoint operator in $l^2$. The space $E_k$ of all $kN$-periodic sequences is
a finite dimensional subspace of $l^\infty$ invariant with respect to $L$. The restriction of $L$ to $E_k$ is
denote by $L_k$.

Let
   $$ Q_k=\left\{n\in\mathbb{Z} \;:\; -\left[\frac{kN}2\right]\le n\le kN-\left[\frac{kN}2\right]-1\right\},$$
where $[x]$ stands for the integer part of $x$. For a sequence $u=\{u_n\}\in l$ we set
$$ R_ku_n=\left\{\begin{array}{lll} u_n & \text{ if } & n\in Q_k,\\
                                    0   & \text{ if } & n\notin Q_k \end{array}\right.$$
and denote by $S_ku_n$ a unique sequence that belongs to $E_k$ and such that
   $$ S_ku_n=u_n\quad \text{if } n\in Q_k.$$
Thus, $R_k$ is a ``{\it cut off\/}'' operator, while $S_k$ is ``{\it periodization}'' operator.  For $u\in E_k$ we
set
   $$ \|u\|_{l^p_k}=\|R_ku\|_{l^p}.$$
For any fixed $k$, $\|\cdot\|_{l^p_k}$, $1\le p\le\infty$, form a family of equivalent norms on $E_k$,
$\|\cdot\|_{l^2_k}=\|\cdot\|_k$ is the standard Euclidean norm on $E_k$ and $(R_ku,R_kv)=(u,v)_k$ is the standard
inner product on $E_k$.

Obviously,
   $$ \|R_ku\|_{l^p}\le\|u\|_{l^p}$$
and
   $$ \|S_ku\|_{l^p_k}\le\|u\|_{l^p}$$
for all $u\in l^p$. The following identities are easy to verify: for every $u\in l^p$, $1\le p<\infty$,
\begin{equation}\label{A.2}
   \lim_{k\to\infty}\|R_ku\|_{l^p}=\lim_{k\to\infty}\|S_ku\|_{l^p_k}=\|u\|_{l^p},
\end{equation}
\begin{equation}\label{A.3}
   \lim_{k\to\infty}\|LR_ku\|_{l^p}=\lim_{k\to\infty}\|R_kLu\|_{l^p_k}=\|Lu\|_{l^p},
\end{equation}
\begin{equation}\label{A.4}
   \lim_{k\to\infty}\|L_kS_ku\|_{l^p_k}=\lim_{k\to\infty}\|S_kLu\|_{l^p_k}=\|Lu\|_{l^p}.
\end{equation}

Let $\lambda\in \mathbb{C}\setminus\sigma(L)$. Then the operator $L-\lambda$ is invertible in $E=l^2$ and for every
$u=\{u_n\}\in l^2$
\begin{equation}\label{A.5}
   (L-\lambda)^{-1}u_n=\sum_{m\in\mathbb{Z}}G(n,m;\lambda)\,u_m,
\end{equation}
where $G(\cdot,\cdot;\lambda)$ is the so-called {\it Green function\/}. Due to the periodicity assumption,
\begin{equation}\label{A.6}
   G(\cdot+N,\cdot+N;\lambda)=G(\cdot,\cdot;\lambda)
\end{equation}
(periodicity of the Green function along the diagonal). Moreover, the Green function possesses an exponential bound
of the form
\begin{equation}\label{A.7}
   \big|G(n,m;\lambda)\big|\le Ce^{-\mu|m-n|},\quad n,m\in\mathbb{Z},
\end{equation}
where $C>0$ and $\mu>0$ can be chosen independent of $\lambda$ as $\lambda$ ranges over any compact subset of
$\mathbb{C}\setminus\sigma(L)$ (see, e.~g. \cite{Teschl}). Representation (\ref{A.5}) and inequality (\ref{A.7})
imply that for every $\lambda\in\mathbb{C}\setminus\sigma(L)$ the operator $(L-\lambda)^{-1}$ is a bounded linear
operator in $l^p$ for every $p\in [1,\infty]$. Moreover, the space $E_k$ is invariant with respect to
$(L-\lambda)^{-1}$ and, for every $u\in E_k$,
  $$ \big\|(L_k-\lambda)^{-1}u\|_{l^p_k}\le C_p\|u\|_{l^p_k},$$
with $C_p>0$ independent of $k$. In particular, $\sigma(L_k)\subset\sigma(L)$.

Let $(\alpha,\beta)$ be a spectral gap of $L$. Denote by $P^+$ and $P^-$ the spectral projectors in $E=l^2$ that
correspond to the parts of $\sigma(L)$ lying in $(-\infty,\alpha]$ and $[\beta,+\infty)$, respectively. Note that
these are orthogonal projectors. The associated spectral subspaces are denoted by $E^+$ and $E^-$, respectively.
Also, we note that $\sigma(L_k)\cap(\alpha,\beta)=\emptyset$. We denote by $P^+_k$ and $P^-_k$ the spectral
projectors in $E_k$ that correspond to $\sigma(L_k)\cap(-\infty,\alpha]$ and $\sigma(L_k)\cap[\beta,+\infty)$,
respectively, and set $E^\pm_k=P^\pm_kE_k$ for the spectral subspaces.

The projectors $P^\pm$ are given by the Riesz formula
   $$ P\pm=-\frac 1{2\pi i}\int_{C_\pm}(L-\lambda)^{-1}d\lambda,$$
where $C_+$ (resp., $C_-$) is a contour in $\mathbb{C}\setminus\sigma(L)$ encircling $\sigma(L)\cap[\beta,+\infty)$
(resp., $\sigma(L)\cap(-\infty,\alpha]$) so that $\sigma(L)\cap(-\infty,\alpha]$ (resp.,
$\sigma(L)\cap[\beta,+\infty)$) lies outside. Therefore, $P^\pm$ is of the form
\begin{equation}\label{A.8}
   P^\pm u_n=\sum_{m\in\mathbb{Z}}K(m,n)u_m,
\end{equation}
where
   $$ K(m,n)=-\frac 1{2\pi i}\int_{C_\pm}G(m,n;\lambda)\,d\lambda.$$
Hence
\begin{equation}\label{A.9}
   \big|K(m,n)\big|\le Ce^{-\mu|m-n|},\quad m,n\in\mathbb{Z},
\end{equation}
with some $C>0$ and $\mu>0$. Note that the right hand part of (\ref{A.8}) is a bounded linear operator in all
spaces $l^p$, $1\le p\le\infty$. Moreover, the operators $P^\pm_k$ are exactly the restrictions of $P^\pm$ from
$l^\infty$ to $E_k$ and, therefore, are given by the same formula (\ref{A.8}). Also it is easy to verify the
operator
   $$ B=(L-\omega)P^\pm=P^\pm(L-\omega)$$
has a representation of the form
\begin{equation}\label{A.10}
   Bu_n=\sum_{m\in\mathbb{Z}}K(m,n)u_m,
\end{equation}
with $K(m,n)$ (not the same as in (\ref{A.8})) satisfying (\ref{A.9}).

\begin{lemma}\label{l.A.1}
   Let $B$  be an operator of the form (\ref{A.10}) satisfying (\ref{A.9}). Then for any $u\in l^2$
      $$ \lim_{k\to\infty}(BS_ku,S_ku)_k=(Bu,u).$$
\end{lemma}

The proof is contained in the proof of Theorem~3, \cite{BrunoPaTv}.

\section{Linking}\label{A.B}
\setcounter{equation}{0} \setcounter{Theorem}{0} \setcounter{lemma}{0} \setcounter{prop}{0}

Here we recall the so-called linking theorem (see \cite{Rabinowitz68,Willem}). Let $X=Y\oplus Z$ be a Banach space
decomposed into the direct sum of two closed subspaces $Y$ and $Z$, with $dim Y<\infty$. Let $\varrho>r>0$ and let
$z\in Z$ be a fixed vector, $\|z\|=1$. Define
   $$ M=\big\{u=y+\lambda z\;:\; y\in Y, \|u\|\le\varrho, \lambda\ge 0\big\}$$
and
   $$ N=\big\{u\in Z\;:\; \|u\|=r\big\}.$$
Denote by $M_0=\partial M$ the boundary of $M$, i.~e.
   $$ M_0=\big\{u=y+\lambda z\;:\; y\in Y, \|u\|=\varrho \text{ and } \lambda\ge 0, \text{ or } \|u\|\le\varrho
                \text{ and } \lambda=0\big\}.$$

Consider a $C^1$ functional $\varphi$ on $E$ and suppose that $\varphi$ satisfies the Palais-Smale condition, i.~e.
any sequence $u^{(j)}\in E$ such that $\varphi\big(u^{(j)}\big)$ is convergent and $\varphi'\big(u^{(j)}\big)\to 0$
contains a convergent subsequence. Suppose also that
\begin{equation}\label{B.1}
   \beta=\inf_{u\in N}\varphi(u)>\alpha=\sup_{u\in M_0}\varphi(u).
\end{equation}
The last assumption means that $\varphi$ possesses the so-called linking geometry.

Let
   $$ \Gamma=\big\{\gamma\in C(M;E)\;:\; \gamma=\mathrm{id} \text{ on } M_0\big\}.$$
Then
   $$c=\inf_{\gamma\in\Gamma}\sup_{u\in M}\varphi\big(\gamma(u)\big)$$
is a critical value of $\varphi$ and
\begin{equation}\label{B.2}
   \beta\le c\le\sup_{u\in M}\varphi(u).
\end{equation}

\end{document}